\documentclass[aps,prb,showpacs,twocolumn,superscriptaddress,tightenlines]{revtex4}

\usepackage[utf8]{inputenc}
\usepackage{t1enc}
\usepackage[icelandic,swedish,english]{babel}
\usepackage{amsmath,amssymb}
\usepackage{amsbsy,pifont}
\usepackage{color}
\usepackage{graphicx}
\usepackage[pdftex,pdfborder=0,breaklinks,colorlinks=false]{hyperref}

\def\figwidth{1}

\begin{document}

\title{Signatures of Wigner Localization in Epitaxially Grown Nanowires}
\author{L.\;H. Kristinsd\'ottir}
\affiliation{Division of Mathematical Physics, Lund University, Box 118, 22100 Lund, Sweden}
\author{J.\;C. Cremon}
\affiliation{Division of Mathematical Physics, Lund University, Box 118, 22100 Lund, Sweden}
\author{H.\;A. Nilsson}
\affiliation{Division of Solid State Physics, Lund University, Box 118, 22100 Lund, Sweden}
\author{H.\;Q. Xu}
\affiliation{Division of Solid State Physics, Lund University, Box 118, 22100 Lund, Sweden}
\author{L. Samuelson}
\affiliation{Division of Solid State Physics, Lund University, Box 118, 22100 Lund, Sweden}
\author{H. Linke}
\affiliation{Division of Solid State Physics, Lund University, Box 118, 22100 Lund, Sweden}
\author{A. Wacker}
\affiliation{Division of Mathematical Physics, Lund University, Box 118, 22100 Lund, Sweden}
\author{S.\;M. Reimann}
\email[Corresponding author, ]{reimann@matfys.lth.se}
\affiliation{Division of Mathematical Physics, Lund University, Box 118, 22100 Lund, Sweden}
\collaboration{Nanometer Structure Consortium, nmC@LU}
\date{December 1, 2010}

\begin{abstract}
It was predicted by Wigner in 1934 that the electron gas will undergo a transition to a crystallized state when its density is very low. Whereas significant progress has been made towards the detection of electronic Wigner states, their clear and direct experimental verification still remains a challenge. Here we address signatures of Wigner molecule formation in the transport properties of InSb nanowire quantum dot systems, where a few electrons may form localized states depending on the size of the dot (i.e. the electron density). By a configuration interaction approach combined with an appropriate transport formalism, we are able to predict the transport properties of these systems, in excellent agreement with experimental data. We identify specific signatures of Wigner state formation, such as the strong suppression of the antiferromagnetic coupling, and are able to detect the onset of Wigner localization, both experimentally and theoretically, by studying different dot sizes.
\end{abstract}

\maketitle

The transition to a Wigner crystal~\cite{wigner1934} can be viewed as a contest between the electronic Coulomb repulsion and the quantum mechanical kinetic energy. If the Coulomb repulsion dominates, the many-particle ground state and its excitations resemble a distribution of classical particles located in a lattice minimizing the Coulomb energy. In the bulk, the transition to a Wigner crystal is only expected for extremely dilute systems \cite{ceperley1980,drummond2004}, while in lower dimensions, or for broken translational invariance, it becomes accessible at higher densities~\cite{tanatar1989,jauregui1993,rapisarda1996}. A lot of work has focused on finite-sized two-dimensional quantum dots~\cite{creffield1999,egger1999,yannouleas1999,filinov2001,reimann2002}, where the crossover from liquid to localized states in the transport properties of the nanostructure has been addressed~\cite{cavaliere2009,EllenbergerPRL2006}. For one-dimensional systems, localization has been reported in cleaved edge overgrowth structures~\cite{AuslaenderScience2005} and for  holes in carbon nanotubes~\cite{deshpande2008}. These highly correlated one-dimensional systems exhibit a variety of fascinating features as reviewed recently~\cite{DeshpandeNature2010}. Here we introduce a third system, based on epitaxially grown semiconductor nanowires, which allows a straightforward application of tunneling spectroscopy compared to the rather involved cleaved edge overgrowth structures and avoids further complications due to the isospin degree of freedom in carbon nanotubes.

\begin{figure}
 \includegraphics[width=\figwidth\columnwidth]{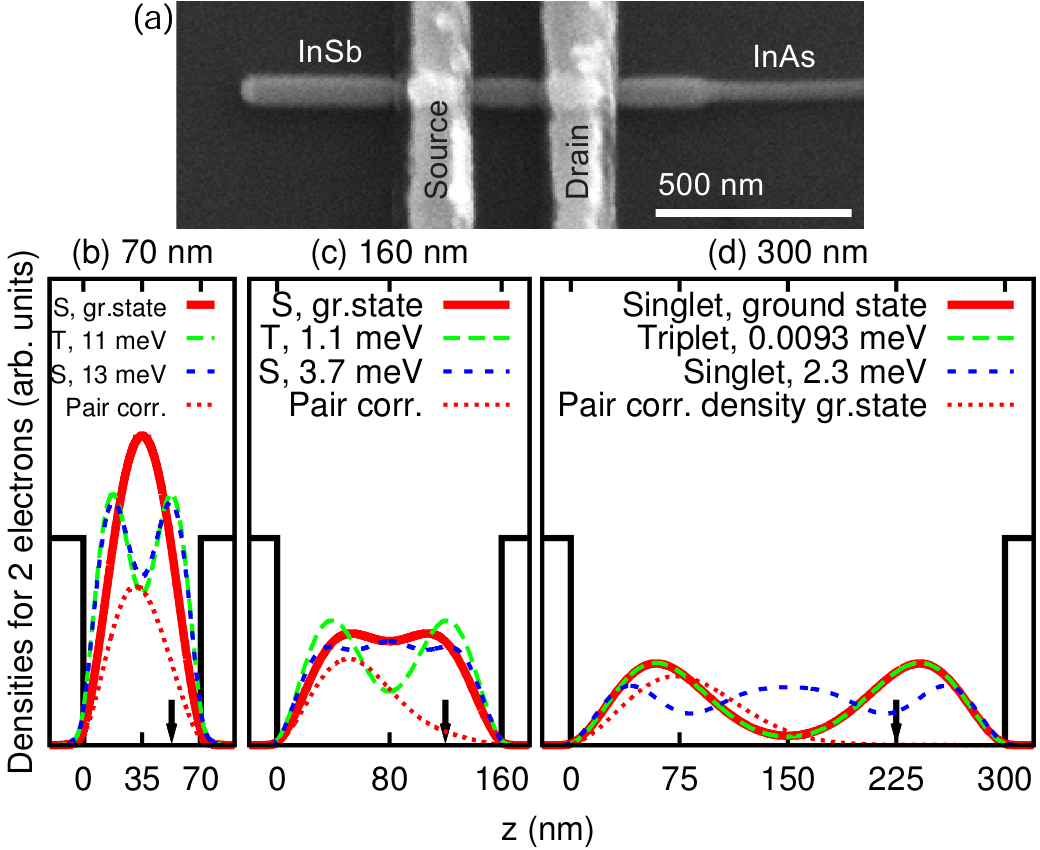}
 \caption{(a) SEM-image of the InSb nanowire on a SiO$_2$ capped Si substrate, where the quantum dot is defined by Schottky barriers of the gold contacts (`source' and `drain'). Calculated electron density in nanowires of lengths 70~nm, 160~nm, and 300~nm is displayed in panels (b,c,d), respectively, for the lowest two-electron states (excitation energies are given; `S' stands for singlet and `T' for triplet). For the two-particle ground state the pair-correlated density is shown with the position of one electron marked by a black arrow.}
 \label{fig:SEMandDensplot}
\end{figure}

InSb nanowires~\cite{NilssonNL2009}, as used here, allow for the realization of quantum dots, where the electronic confinement along the nanowire is established by Schottky barriers to gold contact stripes, see Fig.~\ref{fig:SEMandDensplot}(a). Varying the distance between the stripes (here: 70~nm and 160~nm) allows for the systematic realization of wires with specific length and thereby controlled electron densities.  For our calculations we model the nanowire as a hard-wall cylinder with the experimental radius 35~nm. The Schottky barrier at the semiconductor-metal interface creates a standard quantum well with a width equal to the contact spacing. The Coulomb interaction between the electrons is approximated as that in a cylinder embedded in homogeneous matter, taking into account the different dielectric constants of the wire and the surrounding material~\cite{SlachmuyldersPRB2006,LiPRB2008}. Exact many-particle states in the wire are evaluated with the configuration interaction method. 

The results can be understood in terms of two limiting cases: a short wire with no electron localization and a long wire with Wigner localization\footnote{As we focus on 2-3 electrons, we cannot speak of a macroscopic effect such as Wigner crystallization. Hence the term \textit{Wigner localization}.}.

The first limiting case, where interaction is dominated by kinetic energy, can be described by the independent-particle shell model. There the two-particle ground state is obtained by populating the lowest single-particle level with a spin-up and a spin-down electron. Thus the spatial electron density follows that of the lowest single-particle level and exhibits a peak in the center of the quantum dot. The lowest excited two-particle state is obtained by moving one electron to the first excited single-particle level at the cost of the level spacing energy $\Delta\varepsilon$. Thus one expects the two-particle excitation energy $\Delta E_2\approx \Delta\varepsilon$. Furthermore the spin degrees allow for four realizations of such an excited two-particle state, which are typically split into a triplet and a singlet due to exchange interaction.

In the second limiting case, Wigner localization, the electrons are localized at different positions along the wire, minimizing the Coulomb repulsion. Thus the two-particle ground state density exhibits two peaks and a minimum in the center of the nanowire segment. As the electrons can have arbitrary spin on each site, one has four realizations of this configuration, with a minor energy split between a singlet and a triplet. Hence, we expect a very small $\Delta E_2\ll\Delta\varepsilon$, while further excitations are significantly higher in energy and exhibit a different spatial distribution of charge.

At the onset of localization, the electron density is expected to resemble two weakly separated peaks in the two-particle ground state. The interaction of the electrons is substantial, without yet dominating the kinetic part. Hence the two-particle excitation energy is considerably lower than the single-particle excitation energy, $\Delta E_2<\Delta\varepsilon$. However, as the two electrons are not yet fully crystallized, $\Delta E_2$ is expected to be well above zero.

\begin{figure}
 \includegraphics[width=\figwidth\columnwidth]{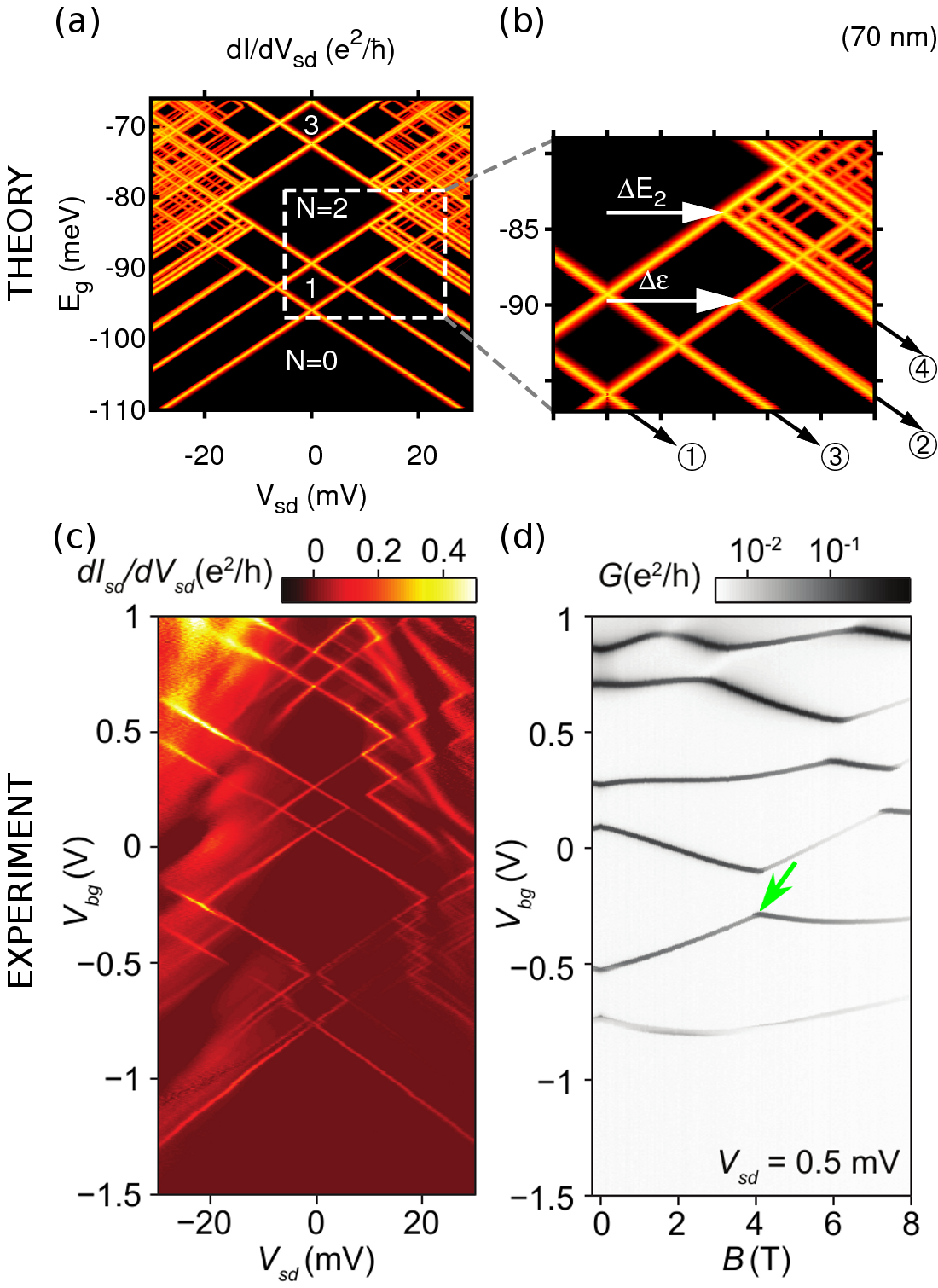}
 \caption{Results for an InSb nanowire of length $L=70$~nm.  (a) Simulated differential conductance as a function of bias ($V_{\text{sd}}$) and gate energy $E_g$. The number of particles in the dot, $N$, is shown in each diamond. (b) A closer look at the area marked by a dashed box in panel (a). The conduction lines, where tunneling into the $N=1$ ground state and first excited state sets in, are marked by the symbols \ding{192} and \ding{193}, respectively. The corresponding lines for the entering of the second electron, where the dot reaches the  $N=2$ ground state and the $N=2$ excited state, are marked by \ding{194} and \ding{195} symbols, respectively. The separation between these lines provides the excitation energies from the $N=1$ and $N=2$ ground states, $\Delta\varepsilon$ and $\Delta E_2$, respectively, which are depicted by arrows. (c) Experimental differential conductance as a function of bias ($V_{\text{sd}}$) and gate voltage ($V_{\text{bg}}$). (d) Experimental differential conductance as a function of magnetic field ($B$) and gate voltage ($V_{\text{bg}}$).}
 \label{fig:L070}
\end{figure}

\begin{figure} 
 \includegraphics[width=\figwidth\columnwidth]{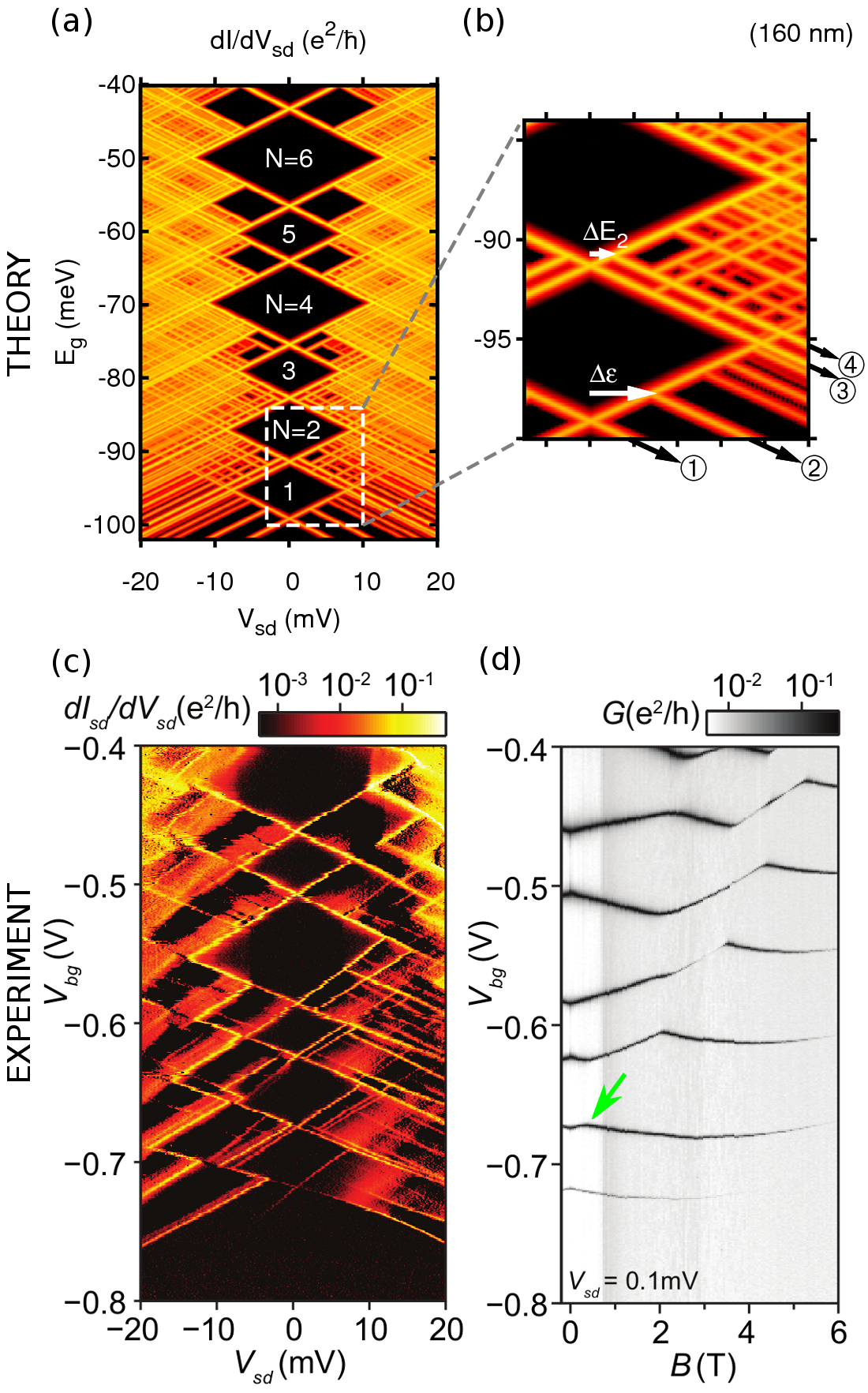}
 \caption{Results for an InSb nanowire of length $L=160$~nm, panels as in Fig.~\ref{fig:L070}}
 \label{fig:L160}
\end{figure}

Tunneling spectroscopy is a convenient way to study ground and excited states in quantum dot systems. Here we can use the gold contacts (Fig.~\ref{fig:SEMandDensplot}(a)) as source and drain by applying a bias $V_\text{sd}$ between both stripes. The nanowire is located on a highly doped Si substrate covered by an insulating SiO$_2$ layer, which allows for application of a back-gate voltage $V_\text{bg}$ providing an approximately homogeneous shift in energy of all levels in the dot. Varying $V_\text{sd}$ and $V_\text{bg}$ provides the characteristic charging diagrams (see e.g. \cite{reimann2002}) displayed in Figs.~\ref{fig:L070}(c) and  \ref{fig:L160}(c) at a temperature of 300~mK. Here high differential conductance indicates that the electron addition energy (affinity) coincides with the chemical potential in either of the gates. The diamonds of vanishing conductance centered around zero $V_\text{sd}$ are the regions of Coulomb blockade, where the chemical potentials of both reservoirs are above the energy difference between the $(N-1)$- and $N$-electron ground state and below the energy difference between the $N$- and $(N+1)$-electron ground state. As no further lines of high conductance are found for lower gate bias, we assume that the lowest diamond corresponds to $N=1$. Half the width of this diamond defines the charging energy $U$.

Based on the calculated many-particle states, electron transport is treated within the master equation model~\cite{ChenPRB1994,KinaretPRB1992,PfannkuchePRL1995} with tunneling matrix elements calculated as in Ref.~\cite{cavaliere2009}. The results are displayed in Figs.~\ref{fig:L070}(a) and  \ref{fig:L160}(a) for the respective experimental samples displayed in panel (c). We find that all Coulomb diamonds agree rather well, which indicates that the radial excitations, which are disregarded in our effectively one-dimensional model, only become of relevance for higher particle numbers in the dot.

Now we focus on the excited states and show, that the experimental conductance data along with our theoretical calculations allow for a verification of the Wigner localization scenario described above.

For a 70~nm wire, the 2-electron density along the wire is a single peak, see Fig.~\ref{fig:SEMandDensplot}(b). This corresponds to the independent-particle shell model as described above. In \autoref{fig:L070}(b) we have marked the lines, where the first electron enters the one-electron ground state and the one-electron excited state, by the symbols \ding{192} and \ding{193}, respectively. This reflects the level spacing $\Delta\varepsilon=12$~meV as shown by the horizontal arrow. Similarly, starting from the one-electron ground state, the second electron enters the dot reaching the two-electron ground state and the two-electron excited state at lines marked by the \ding{194} and \ding{195} symbols. The separation between these two lines represents the excitation energy $\Delta E_2=11$~meV. The four lines, \ding{192}-\ding{195}, can be observed in the experimental data in Fig.~\ref{fig:L070}(c) (this is clearer for negative bias, as the measurement results in the positive bias region most likely suffer from charging of impurity states). From this figure, we read $\Delta E_2^\text{exp}=15~\mathrm{meV}\approx \Delta\varepsilon^\text{exp}=16~\mathrm{meV}$, and hence for the sample of length 70~nm, the experimental data are in good agreement with the independent-particle shell model discussed above.

Note that there is some discrepancy between theory and experiment regarding the value of $\Delta\varepsilon$ and $\Delta E_2$. This could be due to bending of energy levels at the interface of the wire and the gold contacts (Schottky barriers), which makes the wire effectively shorter than the spacing of the contacts. Indeed, simulations of a 60~nm wire give $\Delta\varepsilon=16$~meV and $\Delta E_2=15$~meV.

We can quantify the electron-electron interaction strength by the energy difference between the two-particle ground state and twice the energy of the lowest single-particle level (half-width of the $N=1$ Coulomb diamond). This provides the charging energy $U^\text{exp}=6.5$~meV for the 70~nm sample, as read from Fig.~\ref{fig:L070}(c). That is $U<\Delta\varepsilon$, in accordance with the independent-particle shell model being valid when kinetic energy dominates interaction.

For the 160~nm wire, the 2-electron density in Fig.~\ref{fig:SEMandDensplot}(c) resembles two semi-separated peaks, indicating the onset of Wigner localization (as also seen in the pair-correlated density). In \autoref{fig:L160}, the lines \ding{192}-\ding{195} can be identified both in the simulation and the experiment. The theoretical results give $\Delta E_2=1.0$~meV and $\Delta\varepsilon=2.8$~meV, as in the experiments we observe $\Delta E_2^\text{exp}=1.0~\mathrm{meV}< \Delta\varepsilon^\text{exp}=3.2 \mathrm{meV}$.  Again, this is in agreement with the scenario of onset of Wigner localization discussed above.

Note that if we would neglect the different dielectric consant outside the wire, the onset of Wigner localization would first appear at double the actual wire length. Hence the screening due to the different dielectric constants of the wire and the surrounding material is an important effect and must be included in the modelling.

The energy separation between the singlet and the triplet two-electron state, the antiferromagnetic coupling, can also be manifested by the magnetic field dependence of the differential conductance. The $S_z=1$ part of the triplet is lowered in energy by a magnetic field with respect to the singlet state by $g\mu_B B$, where $\mu_B$ is the Bohr magneton. Fig.~\ref{fig:L160}(d) shows that there is a level crossing at $B_{\rm cross}\approx0.4$~T (marked by an arrow). According to Ref.~\cite{NilssonNL2009} the electronic $g$-factors are around 40 for two electrons in the dot. This provides an energy splitting $\Delta E^{\text{mag}}_2=g\mu_B B_{\text{cross}}\approx1$~meV in full agreement with the calculated value for the 160~nm wire. Note that for the 70~nm wire, the level splitting is no longer linear in the high magnetic field, $B_\text{cross}\approx4$~T, at which the crossing appears (marked by an arrow in Fig.~\ref{fig:L070}(d)). Hence we cannot apply the same method to find $\Delta E^{\text{mag}}_2$ for the 70~nm wire, although its result $\Delta E^{\text{mag}}_2\approx10$~meV is of the correct order of magnitude. The strong suppression of this antiferromagnetic coupling between the two electrons (by an order of magnitude, while changing the length by about a factor of two) is one of the hallmarks of the Wigner crystal state~\cite{DeshpandeNature2010}.

\begin{figure}
 \includegraphics[width=\figwidth\columnwidth]{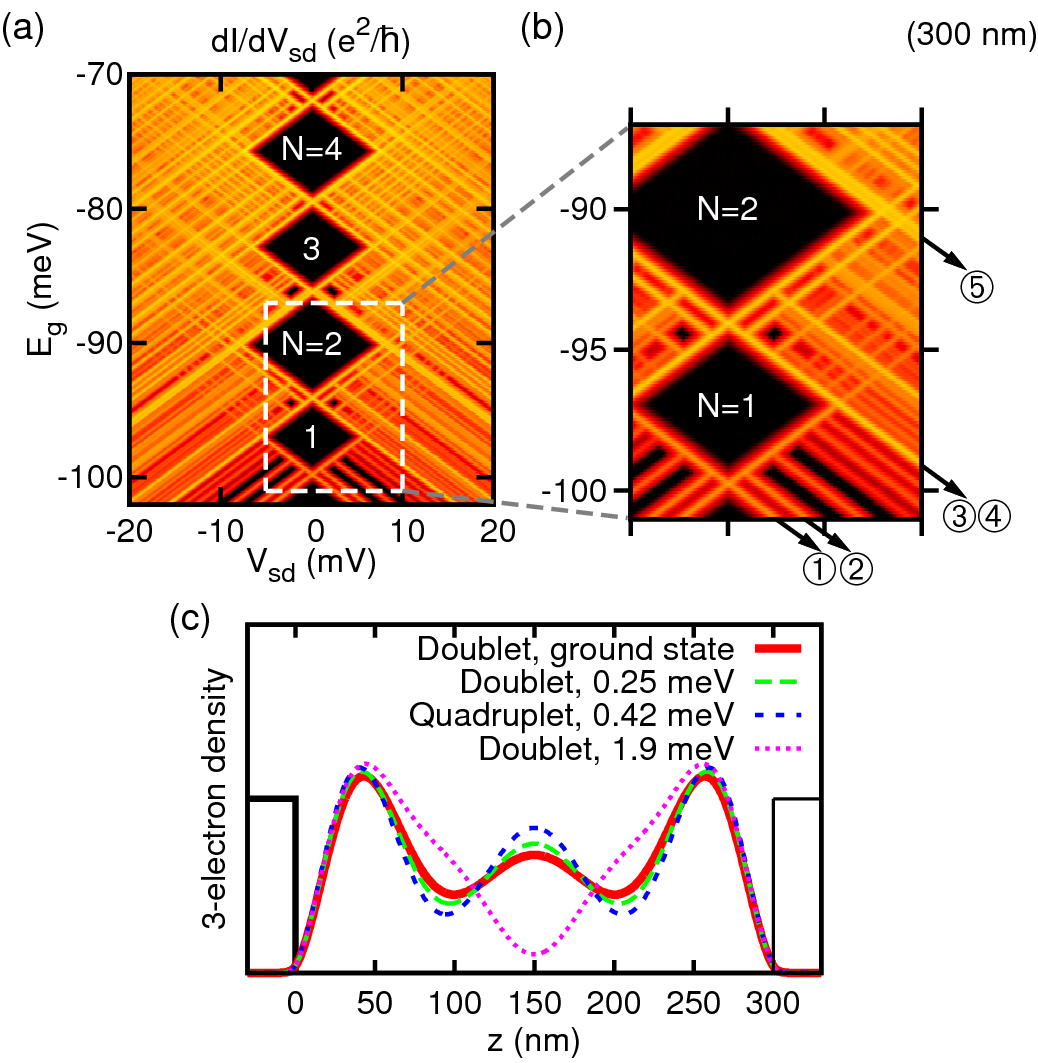}
 \caption{Simulation of a 300~nm long wire. (a) Charge stability diagram. (b) A closer look at the area in the dashed box in panel (a). Symbols \ding{192}-\ding{195} as in Fig.~\ref{fig:L070}. The two lowest $N=2$ states have approximately the same energy, and hence the double conduction line of the 160~nm wire (\ding{194} and \ding{195} in \autoref{fig:L160}b) has merged into a single line leading to the $N=2$ Coulomb diamond. The broad conduction line consisting of three lines for the three lowest $N=3$ states, is marked by the symbol \ding{196}. (c) Electron density of the four lowest $N=3$ states.}
 \label{fig:L300}
\end{figure}

Finally, our theoretical results indicate complete Wigner localization for a 300~nm long wire. Fig.~\ref{fig:SEMandDensplot}(d) shows that in the two-particle ground state, the electrons are strongly localized, i.e. they form a Wigner molecule. From~\autoref{fig:L300}(b) we observe that the conductance line of the $N=2$ triplet first excited state (\ding{194}) has merged into the line of the singlet ground state (\ding{195}), as expected: There is no difference in the energy of these two states, as there should be no difference between the singlet and triplet states of two strongly localized particles. More precisely we find $\Delta E_2=9.3~\mu$eV and $\Delta\varepsilon=0.84$~meV, i.e. $\Delta E_2\ll\Delta\varepsilon$. Furthermore we find $U=5.7$~meV, that is $\Delta\varepsilon \ll U$. This conforms to Wigner localization being present when kinetic energy is strongly dominated by interaction.

Even for the $N=3$ ground state the theoretical calculations suggest the onset of Wigner localization in a 300~nm wire, as seen in~\autoref{fig:L300}(c). The small energy difference between the three lowest $N=3$ states results in a broad conduction line, marked by the symbol \ding{196} in~\autoref{fig:L300}(b). Unfortunately, we could not obtain experimental data for this length, since for such a long sample and low charge densities the effect of disorder is too strong, creating an effective double quantum dot. This can be identified in a charge stability diagram as additional kinks in the conductance lines that comprise the $N=1$ Coulomb diamond~\cite{fuhrer2007}. Such kinks are not present in the stability diagram for the  160~nm wire shown in Fig.~\ref{fig:L160}(c), implying that disorder has no significant effect in that case. Also, Coulomb interaction has been shown to decrease the effect of Anderson localization~\cite{filinov2002}. However the theoretical results demonstrate the prospects of our approach, if more efficient gating schemes are developed.

We have demonstrated the transition from the independent-particle shell model to Wigner localization with increasing length of a semiconductor nanowire sample. While the excitation spectrum follows the independent-particle shell model for the 70~nm wire ($\Delta E_2\approx\Delta\varepsilon$), the onset of Wigner localization is observed for the 160~nm wire ($\Delta E_2<\Delta\varepsilon$) and finally our simulations show complete Wigner localization in a wire of length 300~nm. There the excitation energy of the two-particle state is almost negligible and much lower than the level spacing, $\Delta E_2\ll\Delta\varepsilon$, and the calculated electron density exhibits two peaks.  This shows that InSb nanowires form a convenient system to investigate strongly correlated systems by well established transport measurement techniques.

This work was supported by the Swedish Research Council (VR) as well as the Swedish Foundation for Strategic Research (SSF).


\end{document}